# Diamond Micro-Chip for Quantum Microscopy

*Shahidul Asif [1*], Hang Chen [2*], Johannes Cremer [3,4], Shantam Ravan [3,4,5], Jeyson Támara-Isaza [2,3,6], Saurabh Lamsal[2], Reza Ebadi [3,5], Yan Li [1], Ling-Jie Zhou [7], Cui-Zu Chang [7], John Q. Xiao [2], Amir Yacoby [4,8], Ronald L. Walsworth [3,5,9#], Mark J.H. Ku [1,2#]*

1. Department of Materials Science and Engineering, University of Delaware, Newark, DE 19716, USA

2. Department of Physics and Astronomy, University of Delaware, Newark, DE 19716, USA

3. Quantum Technology Center, University of Maryland, College Park, MD 20742, USA

4. Department of Physics, Harvard University, Cambridge, MA 02138, USA

5. Department of Physics, University of Maryland, College Park, MD 20742, USA

6. Departamento de Física, Universidad Nacional de Colombia, 110911, Bogotá, Colombia

7. Department of Physics, Pennsylvania State University, University Park, Pennsylvania 16802, USA

8. John A. Paulson School of Applied Sciences and Engineering, Harvard University, Cambridge Massachusetts 02138, USA

9. Department of Electrical and Computer Engineering, University of Maryland, College Park, MD 20742, USA

*Equal contribution, # corresponding authors

**Abstract:** The nitrogen vacancy (NV) center in diamond is an increasingly popular quantum sensor for microscopy of electrical current, magnetization, and spins. However, efficient NV-sample integration with a robust, high-quality interface remains an outstanding challenge to realize scalable, high-throughput microscopy. In this work, we characterize a diamond micro-chip (DMC) containing a (111)-oriented NV ensemble; and demonstrate its utility for high-resolution quantum microscopy. We perform strain imaging of the DMC and find minimal detrimental strain variation across a field-of-view of tens of µm. We find good ensemble NV spin coherence and optical properties in the DMC, suitable for sensitive magnetometry. We then use the DMC to demonstrate wide-field microscopy of electrical current, and show that diffraction-limited quantum microscopy can be achieved. We also demonstrate the deterministic transfer of DMCs with multiple materials of interest for next-generation electronics and spintronics. Lastly, we develop a polymer-based technique for DMC placement. This work establishes the DMC's potential to expand the application of NV quantum microscopy in materials, device, geological, biomedical, and chemical sciences.

**Main**

**Introduction**

Nitrogen-vacancy (NV) centers in diamond are a leading modality for sensitive, high-spatial-resolution measurements of magnetic fields. NV-diamond sensors operate under ambient and also extreme conditions (of temperature, pressure, etc.); and have wide-ranging applications in the physical and life sciences, from condensed matter physics and material science [1,2] to chemistry [3], biomedical science [4], geology [5], and circuit analysis [6–8].

One common approach to NV magnetic field imaging is scanning probe microscopy (SPM), typically employing a probe consisting of a single NV near the tip of a diamond nanostructure [9]. SPM spatial-resolution is limited only by the stand-off distance $d$ between the NV and the sample; this enables resolution as small as tens of nm [10]. However, SPM necessarily involves slow pixel-by-pixel scanning, may scratch the sample [11], and is not suitable for liquid or soft matter samples. Furthermore, SPM involves significant cost and complexity, including common NV probe degradation (e.g. picking up dirt or the single NV trapped in the wrong charge state) and the need for high degree of vibration isolation and temperature stability. Another implementation of NV magnetic field imaging is NV wide-field microscopy, using a modality known as the quantum diamond microscope (QDM) [12,13]. Here, a surface-layer of an NV ensemble on a macroscopic diamond crystal is utilized. With a suitable application of microwave (MW) fields to manipulate the NV spins, a measurement of the spatial distribution of the magnetic field across the NV ensemble layer is encoded in the emitted pattern of NV photoluminescence (PL), which is imaged with a camera. The QDM features parallel acquisition of magnetic field information across an entire field-of-view (i.e., no scanning), which enables high-throughput microscopy of both solid and liquid/soft samples [14,15]. Spatial resolution as small as $\sigma \sim 150$ nm [15], limited by optical diffraction, can be achieved. While this resolution is coarser than from NV-SPM, nevertheless sub-µm resolution suffices for a large number of applications. Additionally, there is room for further improvement, e.g. via inhomogeneous ensemble control [15] and machine learning [16]. Given the aforementioned features, the QDM provides a powerful means for high-throughput, high-resolution microscopy for diverse applications [6–8,14,17–22].

In order to achieve diffraction-limited QDM magnetic imaging resolution $\sigma$ of a sample of interest, it is necessary to ensure a small NV layer stand-off distance from the sample $d \ll \sigma$. NVs at a depth ~10 nm can be created reliably without significant detriment to their properties [23], and hence $d$ is primarily determined by how well the diamond contacts the sample. However, creating high-quality diamond-sample interface remains an outstanding challenge that hinders the wider application of the QDM. Depositing the sample material or fabricating the target device directly on the bulk diamond crystal can ensure close proximity between the NVs and the sample, but this approach is possible only for a limited number of materials, e.g., for van der Waals materials that can be placed on diamond via exfoliation or polymer-based transfer [17,20,24–28] or for materials that can be grown on diamond, e.g., via evaporation [29] or sputtering. When working with materials for which this option is not available, a diamond-sample interface has typically been realized by simply placing the bulk diamond crystal (~mm in dimensions) in direct contact with the sample. However, with such a large contact area, there is necessarily non-uniform flatness and particles (e.g., dust) invariably get trapped in between the diamond and the sample, leading to a large and inhomogeneous $d$ across the NV layer that limits the resolution to no better than a few μm [13,19,30]. Furthermore, imaging through diamond, which has a high index of refraction, leads to aberration that can further degrade resolution [19].

A promising approach to these issues is to fabricate a small and thin diamond structure to ensure good diamond-sample interface. For example, diamond nanobeams have been used successfully for single or few NV measurements with nanoscale resolution [31,32]. Here, we report the development and characterization of an optimized diamond micro-chip (DMC) for use in ensemble NV magnetic imaging. Each DMC is tens of μm in lateral dimension and 1-2 μm in thickness; and provides a high-quality interface for the NV sensing layer with both solid-state samples and liquid-phase materials via microfluidics [33,34] (Fig. 1a). In past work, DMCs with lateral dimension ranging from hundreds to tens of μm and thickness less than few μm were used for quantum optics [35–40] and NV sensing applications [7,41–44]. However, several key features of DMCs, necessary for many QDM applications, remain to be demonstrated, including: ensemble NV spin coherence and optical properties similar to that in high-quality bulk diamond; sufficiently uniform strain across the NV layer, to not limit NV sensor performance [17,45–49]; diffraction-limited sub-µm magnetic imaging resolution; a (111)-oriented NV ensemble, to enable use of a vertical bias magnetic field; and reliable, deterministic placement of the DMC on the sample of interest. In this work, we realize DMCs with all these desired properties; outline their deposition and manipulation; characterize their properties; and, as an example DMC application, perform sensitive wide-field magnetic imaging of electrical current. We also discuss possible future uses of DMCs.

## **Results**

DMC fabrication and positioning begins with a protocol reported in Refs. [37–39]. An array of DMCs are created with lateral dimensions 10×10, 20×20, and 50×50 µm$^2$, and thickness ~1-2 µm. Each chip is attached to a diamond mainframe via a small bridge. Nitrogen implantation at an energy of 10 keV with appropriate dosage creates an ensemble NV surface layer with a density [NV] ~1000 µm$^{-2}$ and average depth ~10 nm. DMCs are fabricated from single-crystal CVD diamond substrate with (111) orientation, which leads to one class of NVs with quantization axis, set by the substitutional nitrogen and the vacancy, along the vertical ($z$) direction.

To detach a single DMC and subsequently position it at a target location, we employ a sharp tip on a micro-positioner (Fig. 1b-d). The diamond mainframe containing the DMC array, with the NV surface facing down, is attached to a 3D translational stage and a tilt stage, which enables one to bring the DMC array close to the substrate; this ensures the NV surface remains face down when the desired single DMC is detached. When this DMC is broken off, part of the bridge remains on the chip. The off-center position of the bridge introduces chirality, and once a DMC has landed at the target location, one can confirm that the NV surface is down by looking at the location of the bridge and comparing it to the image of the chip prior to detachment (Fig. 1e). We used this procedure to show that single DMCs with ensemble (111) NV layers can be integrated with a diverse range of materials of interest for next generation electronics and spintronics (Fig. 1e-i): yttrium iron garnet grown via liquid phase epitaxy, a magnetic insulator with low magnon damping; magnetically-doped topological insulator hosting quantum anomalous Hall effects [50,51], grown via molecular beam epitaxy; altermagnet $RuO_2$ [52–54] Hall-bar device; a thin flake (<10 nm thick) of 50% Co-substituted $Fe_5GeTe_2$, which is a room-temperature van der Waals antiferromagnet [55] ; and a superconducting Josephson junction, often used as a qubit building block.

In ensemble NV magnetometry — using optically detect magnetic resonance (ODMR) and related NV measurement techniques— one typically applies an external bias magnetic field $B_0$ along a particular NV axis so to be optimally sensitive to the component of the sample magnetic field $B_z$ parallel to this axis. (A discussion of the NV spin Hamiltonian, ensemble NV measurement techniques, and the effects of strain and off-axis magnetic fields is given in [56]). However, ensemble NV magnetometry to date has predominantly been performed with diamonds having (100) crystal orientation, including work using DMCs [7,41–44]. (100) diamond has all four NV axes oriented at 54.7° with respect to the vertical, which is typically inconvenient and/or non-optimal for many applications, particularly in materials science [57]. In this work, we solve this problem by using DMCs fabricated from (111) orientation diamond, with one NV axis along the vertical direction.

NV-diamond sample fabrication can lead to modification of the local strain environment [17,45,47], with inhomogeneous strain being detrimental to NV magnetometry [46,48,58]. Thus we characterize the spatial profile of the effective strain field $M_z(x,y)$ in example DMCs by QDM strain microscopy of the (111)-oriented NVs, using the technique described in Refs. [45,46,48] (further details described in Supplementary Materials). In Fig. 2, we show the measured $M_z$ profile of two characteristic 10×10 µm DMCs. Two observations can be made immediately. First, the variation of $M_z$ is generally below 1 MHz within the bulk of the DMC. Furthermore, a dramatic strain profile associated with lattice dislocations, such as observed in Ref. [46], is absent. To provide a more quantitative assessment of the strain characteristics, we examine the statistics of the measured $M_z$. A histogram of $M_z$ (Fig. 2c,d) shows a mono-modal distribution with a spread approximately given by the standard deviation $\Delta M_z$. The mean is $\overline{M_z}$=-0.2 and -0.55 MHz for the two DMCs. The differing values may arise from various sources during diamond CVD growth. What is more important for ensemble NV sensing is that the spread of the measured $M_z$ is small in each DMC; i.e., there is an overall, homogeneous strain environment in each chip. The standard deviation is $\Delta M_z$=0.1 and 0.033 MHz for the two DMCs, corresponding to a fractional strain of $3\times10^{-6}$ and $1\times10^{-6}$, respectively, and comparable to the state-of-the-art value of $1\times10^{-6}$ found in the quiet area of a bulk diamond single- crystal [59]. This result demonstrates that fabricated DMC has favorable strain profile not too different from that of a bulk single-crystal diamond.

Beyond ensemble NV magnetometry and QDM magnetic imaging, these results have implications for a proposed approach [59] to use strain microscopy in diamond for directional detection of weakly interacting massive particles (WIMPs), a candidate form of dark matter (DM). (Detailed discussion in Supplementary Information). To reconstruct possible WIMP-induced strain tracks with ~10 nm resolution while avoiding aberration, a typical mm-scale diamond needs to be thinned such that the WIMP impact site is close to the surface (within ~1 µm). A key feasibility question is whether this fabrication process may introduce strain features that mask the strain signal induced by

WIMPs. The standard deviation of strain measured in this work is ~1-3×$10^{-6}$, comparable to the expected WIMP signal ~$10^{-6}$ [59]. Hence, the present DMC results show that it is possible to fabricate a thin diamond structure and be able to maintain sufficiently low strain variation to retain sensitivity to WIMPs, motivating future work to characterize background strain features at the nanoscale.

Next, we characterize the spin coherence properties of the (111)-oriented NVs in an example DMC. Fig. 3a shows a measured NV ODMR spectrum, with a resolved hyperfine-splitting due to $^{14}$N nuclear spins; and a linewidth (half-width half-maximum) of about 1.2 MHz that is comparable to NVs in bulk diamond generated at a similar density and implantation energy [23]. For the same DMC, we also measure an ensemble NV Hahn-echo spin coherence time $T_2 \approx 2.3$ μs (Fig. 3b) and a longitudinal spin relaxation time $T_1 \approx 0.46$ ms (Fig. 3c). For shallow NVs (~ tens of nm), $T_2$ and $T_1$ decrease rapidly with NV depth $d$. For a standard (100)-oriented diamond, simulations using Stopping Range of Ions in Matter (SRIM) [60] estimate an average stopping range of the nitrogen ion to be ~15 nm for an implantation energy of 10 keV [61,62]. However, for (111)-oriented diamond, SRIM finds that the stopping range can be reduced by as much as a factor of 2 [63]; hence, $d$ ~ 10 nm or below may be expected for NVs in the present DMC. To estimate $d$, we refer to the measurement performed in Ref. [64], which reports $T_1$ and $T_2$ of several shallow NVs with experimentally characterized $d$. Using these results, a typical NV depth $d$~8 nm best corresponds to the measured combination of NV $T_1$~0.5 ms and $T_2$~2 μs in our DMC. This value is consistent with the measurement reported in Ref. [23], where $T_2$~2.5 μs is found for an NV ensemble with $d$ ~10 nm. Hence, we conclude that the ensemble NV spin coherence properties in the present DMC are consistent with those at a similar depth and density in standard bulk diamond crystal.

As an application demonstration, we perform high-resolution wide-field NV magnetic imaging using a 20×20 µm DMC deposited on a fabricated meandering wire phantom, through which we send a DC current to generate a target magnetic field (Fig. 4a). The (111)-oriented DMC allows us to directly measure $B_z$, the $z$-component of the phantom magnetic field. Fig. 4b shows a simulated $B_z$ map, based on a simple model of current flow through the phantom. The measured $B_z$ map is shown in Fig. 4c, in reasonable agreement with the simulation, other than for a small area around the corners where complexity in the real current flow is likely present and not captured by the simple model.

To characterize the spatial resolution σ of ensemble NV fluorescence imaging using a DMC, we extract the minimum observable size of small features in the DMC strain map of $M_z$. Fig. 4d shows an example, with two strain features in close proximity. We fit $M_z$ to two Gaussian profiles and determine the size of the smaller feature to be 450(30) nm, consistent with the expected diffraction-limited spatial resolution of about 440 nm (Supplementary Materials).

A key to DMC utility is to be able to conveniently and reliably place the chip at the desired location relative to the target sample. When a DMC detaches, it is not possible to control precisely where it lands; hence, most often it will be some distance away from the target. The common approach is then to push or drag the DMC to the desired location using a sharp tip [7,37–39,42–44]. However, this technique has a high probability of damaging the sample or rendering the tip and/or DMC unusable (e.g., by picking up dirt or the DMC becoming stuck). Thus, the present fully-tip-based method is non-optimal for high-throughput DMC applications.

Inspired by van der Waals (vdW) material assembly, we employ a polymer-assisted dry-transfer process for DMC manipulation, summarized in Fig. 5(a)-(e) (details discussed in Supplementary

Materials). With a sharp tip, we first break a DMC off from the diamond mainframe onto a substrate (this may be a separate substrate from the target sample). We then use a patterned polymer microstructure on a transparent thin glass slide to pick up the DMC and bring it to the desired location on the target sample. To release the DMC, we tilt the slide so that the chip is gradually detached from the polymer micro-structure. Lastly, a tip can be used for fine-tuning of DMC placement; as there is no need for the DMC to traverse a long distance, little movement is required for the tip and hence there is minimal risk to the sample, tip, and DMC. In Fig. 5(f)-(l), we show the sequence applied to the deposition of a 50×50 μm$^2$ DMC on a RuO$_2$ altermagnet device with Ti/Au leads. We note that recently, Ref. [41] also demonstrated a different dry-transfer technique to place a similarly-sized DMC on vdW materials. Compared to that work, our technique has the following features (see Supplementary Materials for further details): (i) there is no need to heat the sample, which is desirable for temperature- or air-sensitive materials; (ii) no additional fabrication procedure is needed; and (iii) instead of being permanently attached to the target sample, the DMC can be reused. For example, the DMC in Fig. 5(l) was subsequently picked up again and transferred to another RuO$_2$ device (Fig. 5(m)). Moreover, in Fig. 5(n)-(p), we show that two DMCs, already placed on a permalloy spin-torque device, can be picked up again using a polymer microstructure and positioned on new devices; this result shows that we can selectively pick up an individual DMC from a collection of nearly-spaced chips. This technique enables a reliable, high-throughput process where one first breaks off all the DMCs onto a substrate for storage; and then one simply transfers an individual DMC from the storage substrate to the target sample, repeating the process for other DMCs and possibly other samples, as desired.

## Conclusion

In conclusion, we characterize a diamond micro-chip (DMC) with a (111)-oriented NV ensemble in a nanoscale surface layer, and demonstrate its utility for high-resolution quantum microscopy. We find that the DMC has a suitably homogeneous strain profile and ensemble NV spin coherence properties consistent with NVs in bulk diamond at a similar depth from the surface. We perform high-resolution wide-field NV magnetic imaging using a DMC deposited on a wire phantom carrying a DC current; and show that diffraction-limited resolution can be achieved for ensemble NV fluorescence imaging using a DMC. We also demonstrate the ability to interface a DMC with a diverse set of materials, and discuss implications for diverse applications. Lastly, we develop a polymer-based pick-and-place technique for reliable DMC manipulation and localization. Our work reveals the potential of DMCs with an (111)-oriented NV ensemble to expand the application of NV quantum sensing in wide-ranging fields such as materials, device, geological, biomedical, and chemical sciences.

## Acknowledgements

This work was supported by NSF award 2203829; the University of Delaware Research Foundation – Strategic Initiative award; the Army Research Laboratory MAQP program under Contract No. W911NF-19–2–0181; and the University of Maryland Quantum Technology Center. This research was partially supported by NSF through the University of Delaware Materials Research Science and Engineering Center DMR-2011824 Seed Award program. JTI acknowledges support from the Universidad Nacional de Colombia, project No. 57522. The quantum anomalous Hall insulator growth and fabrication done at Penn State is supported by the ARO Award (W911NF2210159). C. Z. C. acknowledges the support from Gordon and Betty

Moore Foundation's EPiQS Initiative (GBMF9063 to C. Z. C.). AY acknowledges support from the Army Research Office under Grant Number: W911NF-22-1-0248, the Gordon and Betty Moore Foundation through Grant GBMF 9468, and by the Quantum Science Center (QSC), a National Quantum Information Science Research Center of the U.S. Department of Energy (DOE). We thank Xi Wang (University of Delaware) and Philip Kim (Harvard University) for providing probe stations for diamond micro-chip deposition. We thank Andrew F. May and Michael A. McGuire for providing the Co-substituted $Fe_5GeTe_2$ crystals, Nikola Maksimovic for fabricating the Josephson junction, Xinhao Wang for growing $RuO_2$, and Tristan Timog and Matthew Coughlin for experimental support.

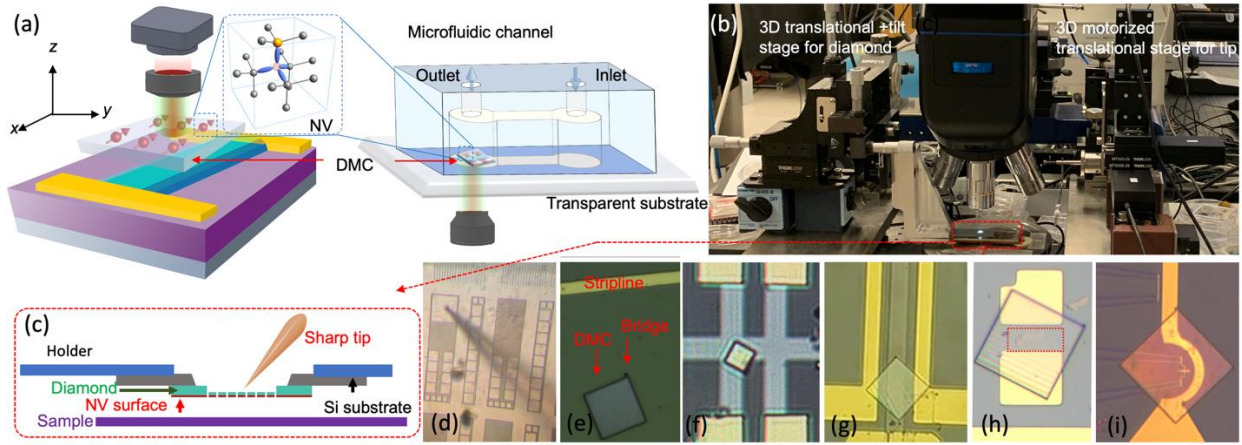

*Figure 1. (a) A diamond micro-chip (DMC), with tens of μm lateral dimension and 1-2 μm thickness, enables a high-quality ensemble NV-sample interface. A DMC contains an ensemble of NVs in a surface layer (tens of nm thick), which enables wide-field magnetic imaging with a camera — a modality known as the quantum diamond microscope (QDM). Each DMC employed in this work contains (111)-oriented NVs (inset), which allows the application of an out-of-plane bias magnetic field $B_0$ and measurement of an out-of-plane component of the sample magnetic field $B_z$. A DMC can also be conveniently integrated with a microfluidic device. (b, c, d) Schematics for depositing a single DMC in target location. Diamond sample containing arrays of DMCs is mounted to a Si substrate attached to a holder. The holder is manipulated with a 3D translational and tilt stage, which serves to bring the desired DMC array into the field of view of a microscope and above the location where a single DMC is to be deposited. A sharp tip controlled by a 3D motorized translational stage is used to detach the desired DMC and subsequently nudges it to the target location. (b) Deposition apparatus with a microscope, stages for the diamond and tip, and holder (diamond and sample is underneath the holder). (c) Cartoon schematic of the deposition process. (d) Microscope image of a tip near a DMC. An array of DMCs with sizes 50✕50, 20✕20, and 10✕10 μm² is visible; each DMC is attached to a diamond mainframe via a*

*narrow bridge. A sharp tip detaches the desired single DMC and then subsequently nudges this DMC into the target location. (e) The bridge that previously connected a DMC with the diamond mainframe introduces chirality and enables one to confirm that the NV side is indeed facing downward. (e-i) DMC can be deposited on a diverse range of materials of interest for next-generation electronics and spintronics: (e) yttrium iron garnet (YIG) thin-film next to an MW stripline; (f) magnetically-doped topological insulator Hall-bar device; (g) altermagnet $RuO_2$ Hall-bar device; (h) <10 nm thick flake (flake highlighted by the dashed rectangle) of 50% Co-substituted $Fe_5GeTe_2$ , a room-temperature van der Waals antiferromagnet; and (i) a Josephson junction.*

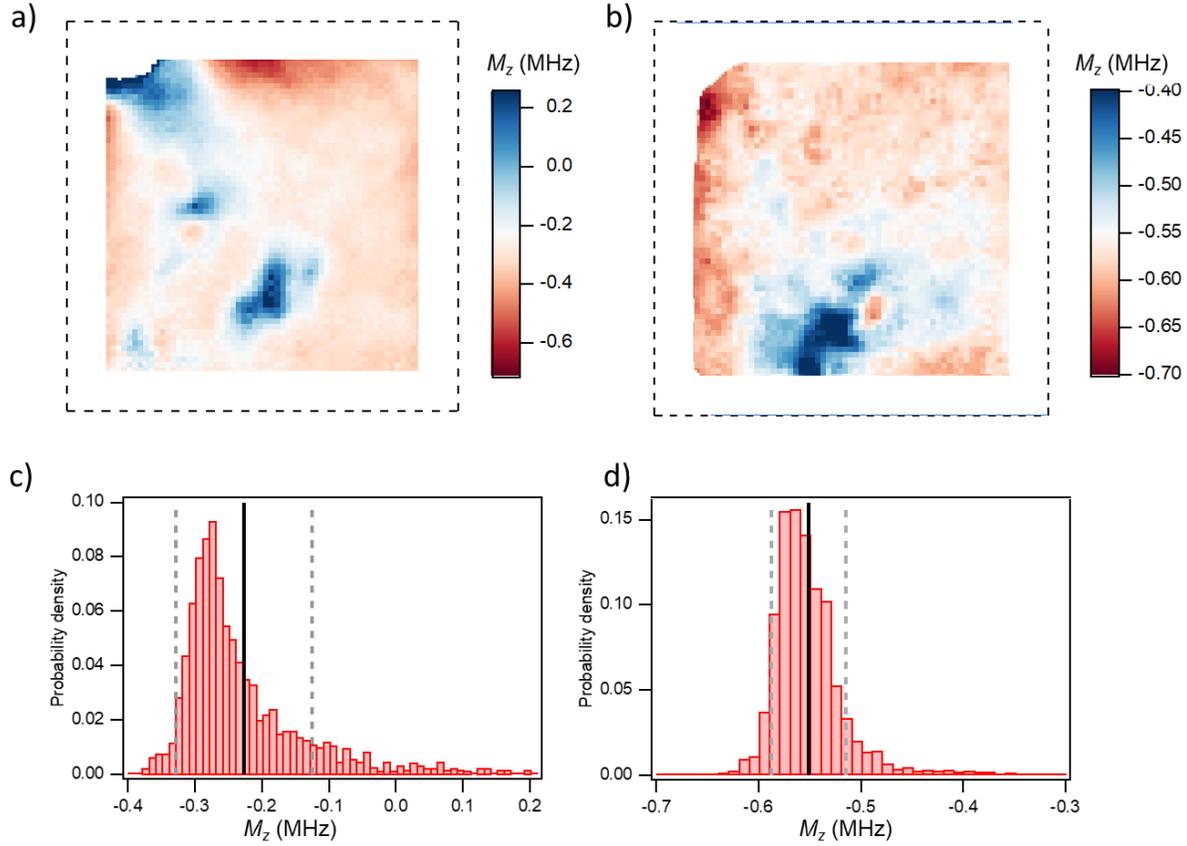

*Figure 2. Measured distribution of axial strain field $M_z$ reveals variation of local strain experienced by (111)-oriented NVs in DMC. (a) and (b) $M_z$ maps from two different DMCs with lateral dimensions of 10×10 μm. Dashed edges show DMC boundaries. Strain image pixel size is 0.14 μm. (c) and (d) Histograms of $M_z$ values from DMCs shown in (a) and (b), respectively. Mean $\overline{M_z}$ and standard deviation $\Delta M_z$ are $\overline{M_z}$=-0.2 MHz, $\Delta M_z$=0.1 MHz for the DMC shown in (a), and $\overline{M_z}$=-0.55 MHz, $\Delta M_z$=0.033 MHz for the DMC shown in (b). Mean and one standard deviation values are shown in the histograms as solid and dashed lines, respectively.*

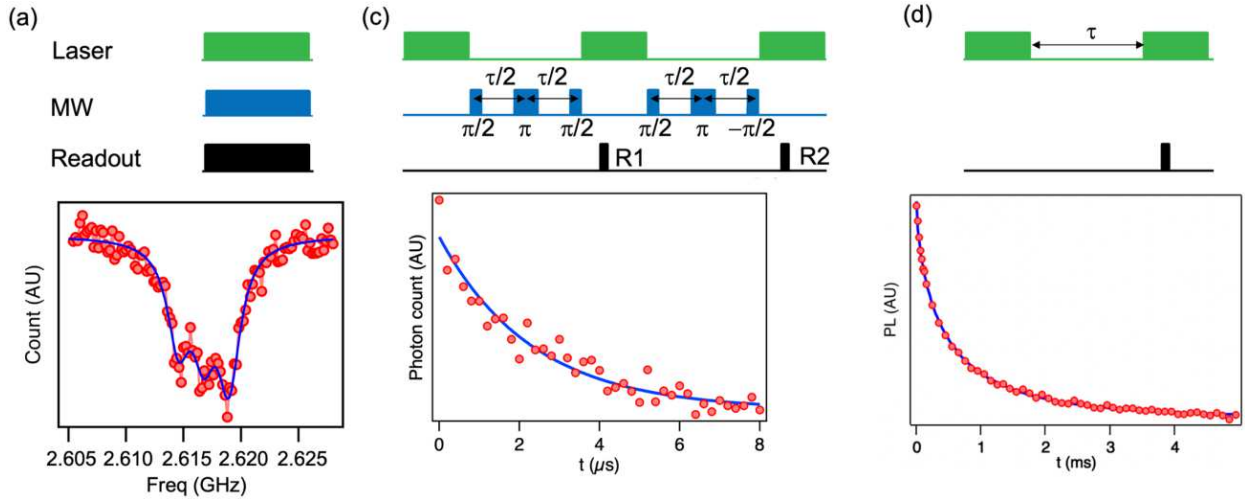

*Figure 3. Spin coherence properties of (111)-oriented NVs in DMCs. The schematics of measurement sequence is shown in the top of each panel. (a) Optically detected magnetic resonance showing that hyperfine splitting is resolved and linewidth (half width half maximum) is 1.18(6) MHz. (b) Hahn spin-echo measures $T_2 \approx 2.3(3)$ μs. (c) Relaxometry measurement obtains a longitudinal relaxation time $T_1 \approx 0.46(4)$ ms.*

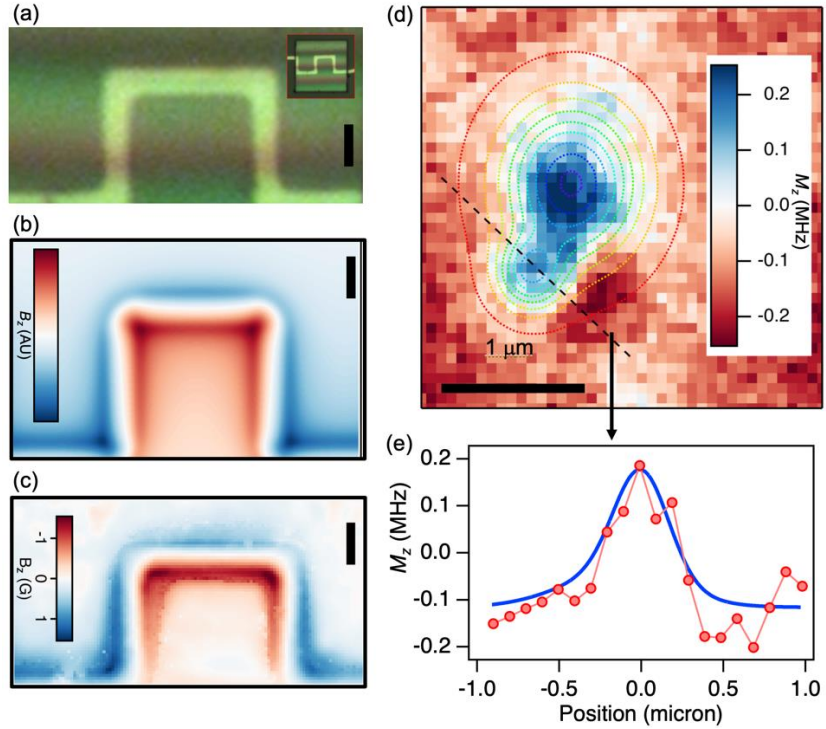

*Figure 4. Demonstration of wide-field magnetic imaging using (111)-oriented NVs in a 20×20 μm DMC. (a) Optical image of a meandering phantom as seen through a DMC. Inset: DMC is deposited on a fabricated meandering wire phantom. (b) Simulated vertical (z) component of magnetic field $B_z$ generated by DC current in phantom. (c) Ensemble NV map of $B_z$ measured with current in phantom. The overall pattern matches that of simulation in panel (b), with minor differences due to complexity in current flow around wire corners. Scale bar corresponds to 5 μm for (a-c). (d) Two localized features in axial strain field $M_z$ used to benchmark spatial resolution of ensemble NV magnetic imaging using a DMC. $M_z$ is fit to two Gaussians; contour of the two-dimensional fit is shown. (e) Linecut of $M_z$ data (fit) along dashed line in (d) shown as red points (blue curve). Extracted spatial resolution is 450(30) nm.*

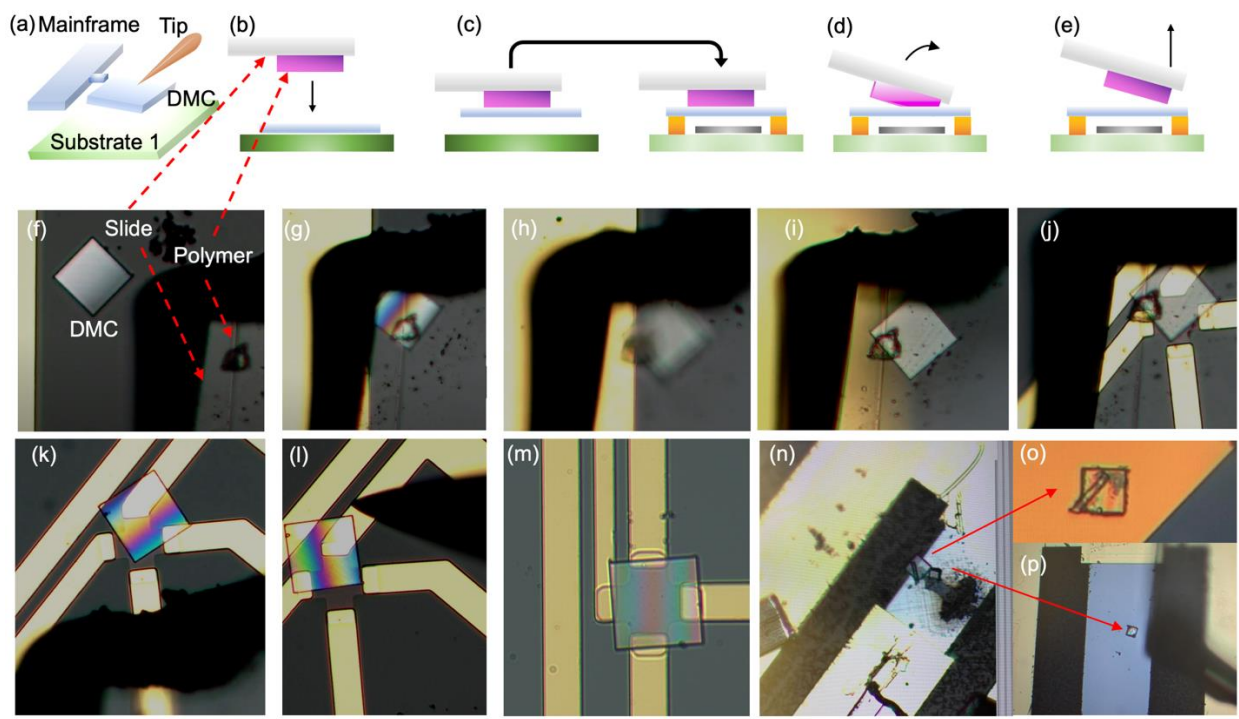

*Figure 5. Polymer-assisted dry-transfer technique for efficient DMC manipulation and placement. (a-e) Schematics. (a) A DMC is first broken off from the diamond mainframe (e.g., with a tip) and lands on a substrate. (b) A transparent substrate (e.g., a glass slide) with a polymer microstructure approaches the DMC. (c) The polymer assembly picks up the DMC and places it on a sample of interest. (d) Tilting the slide releases the DMC from the polymer. A docking structure with adhesive property can also further facilitate the release. (e) The polymer assembly is pulled off, leaving the DMC on the sample. (f-l) Optical images of the polymer DMC deposition process. (f) A DMC, after being broken off, sits on a substrate. In the image one also sees a glass slide with a polymer microstructure. (g) The polymer assembly approaches and comes in contact with the DMC. (h-i) The polymer assembly picks up the DMC. (h) shows the optical image when the substrate is still in focus; the DMC is out of focus, showing that it is picked up. (i) shows the DMC in focus, whereas the substrate is no longer in focus. (j) The polymer assembly brings the DMC on top of a $RuO_2$ altermagnet Hall-bar device and deposits the DMC on the sample. (k) The polymer assembly exits,*

*while the DMC remains on the sample. (l) A tip can be used to further fine-tune the DMC position. (m) The DMC from (l) is picked up again by polymer assembly and transferred to another $RuO_2$ Hall-bar device. (n-p) The technique can also selectively transfer an individual DMC from a collection of closely-spaced chips. Among the two DMCs on a permalloy spin-torque device (n), one is transferred to an Au structure (o) and another transferred to a new spin-torque device (p).*

# Supporting Information: Diamond Micro-Chip for Quantum Microscopy

*Shahidul Asif [1\*], Hang Chen [2\*], Johannes Cremer [3,4], Shantam Ravan [3,4,5], Jeyson Támara-Isaza [2,3,6], Saurabh Lamsal[2], Reza Ebadi [3,5], Yan Li [1], Ling-Jie Zhou [7], Cui-Zu Chang [7], John Q. Xiao [2], Amir Yacoby [4,8], Ronald L. Walsworth [3,5,9#], Mark J.H. Ku [1,2#]*

1. Department of Materials Science and Engineering, University of Delaware, Newark, DE 19716, USA

2. Department of Physics and Astronomy, University of Delaware, Newark, DE 19716, USA

3. Quantum Technology Center, University of Maryland, College Park, MD 20742, USA

4. Department of Physics, Harvard University, Cambridge, MA 02138, USA

5. Department of Physics, University of Maryland, College Park, MD 20742, USA

6. Departamento de Física, Universidad Nacional de Colombia, 110911, Bogotá, Colombia

7. Department of Physics, Pennsylvania State University, University Park, Pennsylvania 16802, USA

8. John A. Paulson School of Applied Sciences and Engineering, Harvard University, Cambridge Massachusetts 02138, USA

9. Department of Electrical and Computer Engineering, University of Maryland, College Park, MD 20742, USA

*Equal contribution, # corresponding authors

## Section 1. Robustness of DMC-sample interface

Once a DMC is properly placed on a sample, the sample can be transported between different locations and the DMC remains in place without being disturbed. A DMC also remain in its place after a wire bonding process. We note that it is generally necessary to deposit the DMC first before wire bonding, as the bonding wires will be in the way for the diamond sample containing DMC array to approach the sample of interest. Likewise, polymer-assisted transfer should generally take place prior to wire bonding. Lastly, we note that DMCs can also be held against gravity even without specialized surface preparation or adhesive material such as employed in Refs. [1,2]. For example, we have shown that with DMCs deposited on YIG or quartz substrates, when we flip the substrate over (e.g., in a measurement setup where the sample has to face down), the DMCs remain on the sample.

## Section 2. Experimental setup.

The setup for wide-field microscopy of strain (Fig. 2 of the main text) and stray magnetic field (Fig. 4c of the main text) is described previously [3,4]. A 532 nm laser (Coherent Verdi 2G) provides excitation. The light is sent through a Kohler-illumination system consisting of a beam expansion lens and a 0.9 NA air objective, which expands the beam to illuminate an area of up to ~ 50×50 $\mu m^2$ on the sample. NV photoluminescence (PL) is collected by the same objective. The collected light is separated from the excitation via a 552 nm edge dichroic (Semrock LM01-552-25). Note that the beam expansion lens is located between the laser and the dichroic; the collected light does not pass through this lens. After the dichroic, another 570 nm long-pass filter (Semrock BLP01-647R-25) further removes light not in the PL wavelength range. The collected light is imaged via a tube lens (focal length f = 200 mm) onto a CMOS camera (Basler acA1920-155um). 2 × 2 binning is applied during data analysis, and each pixel corresponds to 140 nm on the sample. A microwave (MW) signal is sourced by a signal generator (Stanford Research Systems SG384 or Windfreak USB3) and an amplifier (Minicircuits ZHL-16W-43-S+). The sample is fixed on a printed circuit board (PCB) with a co-planar waveguide (CPW). The sample substrate has a stripline that is wirebonded to the CPW of the PCB, to which MW is delivered. The MW circuits ends with an MW terminator. For stray-field microscopy, the CPW also serves to provide the electrical connection between the phantom wire and the current source. The sample and the PCB are mounted on a combined manual translation stage and piezo positioner (Thorlabs NanoMax or PhysikInstrumente P-615.3CD NanoCube XYZ Piezo System). A permanent magnet mounted on a three-axes translational stage is used to provide an external bias magnetic field. Generally, the magnet is placed right beneath the sample to provide a vertical field aligned with (111)-oriented NVs.

The same system is modified as described below for pulsed measurements (spin echo and spin relaxation in Fig. 3 of the main text). Excitation light is modulated with an acousto-optical modulator (AOM) powered by a radiofrequency (RF) driver (Gouch & Housego AOMO 3250-220 532nm and AODR 1250AFP-AD-6.6). Beam expansion light is removed so the excitation is focused onto a tight spot on the sample. The collected PL, instead of going through a tube lens and then imaged onto a camera, is directed into a single-mode fiber for confocal microscopy.

The PL is finally detected by a single photon counting module (Excelitas 28252). Photon counting is performed with a Data Acquisition System (National Instrument USB-6361, X Series DAQ Device with BNC Termination). A MW signal is supplied by Stanford Research Systems SG384 with built-in IQ control, and modulated with a switch. The NV interrogation pulse sequence is generated by a Swabian Instrument Pulse Streamer 8/2.

**Section 3. Strain and stray-field microscopy.**

Wide-field microscopy is performed following the technique described in Refs. [3–7]. We align an external bias magnetic field with (111)-oriented NVs. We perform optically-detected magnetic resonance (ODMR) measurements by continuously illuminating the NVs with the excitation light while driving the NVs with an MW field. The MW frequency $\nu$ is swept, and an ODMR image is captured at each MW frequency. To improve the signal-to-noise ratio (SNR), this sequence is repeated multiple times to create a set of images $PL_{i,\nu}(x,y)$. We perform post-measurement alignment of images to counteract the effect of drift [3], before summing the images at a given $\nu$ to create a signal-averaged image $PL_\nu(x,y)$. Then, at a given pixel $(x,y)$, we create an ODMR spectrum PL vs $\nu$, and fit the spectrum to extract the resonance position. This procedure is performed to extract a spatial map of both the upper and lower NV resonances, $\nu_\pm(x,y)$, corresponding to the the $m_s = 0 \leftrightarrow \pm 1$ transitions.

A simplified Hamiltonian for the NV electronic spin (assuming (111)-orientation) is

$$H = (D_g + M_z)S_z^2 + \gamma_e \mathbf{B} \cdot \mathbf{S},$$

where $D_g$=2.87 GHz is the ground-state zero-field splitting, $M_z$ is the component of the effective electric field associated with strain (which we will call effective strain field) coupling to the $S_z^2$ term, $\gamma_e$=28 GHz/T is the NV gyromagnetic ratio, **B** is the net magnetic field applied to the NV, and **S**=($S_x$, $S_y$, $S_z$) is the $S$=1 spin operator. Here, the $z$-axis corresponds to the quantization axis set by the nitrogen and vacancy, which is also the vertical axis for (111)-oriented NVs. In writing down this Hamiltonian, we ignore higher-order contributions from strain, which are negligible in the context of magnetometry. To leading order, the spin resonance corresponding to the $m_s = 0 \leftrightarrow \pm 1$ transitions are $\nu_\pm = D \pm \gamma_e B_z + 3\gamma_e^2 B_\perp^2/(2D)$, where we have defined the effective zero-field splitting (ZFS) $D \equiv D_g + M_z$ and $B_\perp$ is the off-axis field.

For strain microscopy (Fig. 2), we obtain the spatial map of effective zero-field splitting $D \equiv D_g + M_z$ by extracting the common-mode shift, $D(x,y)=[\nu_+(x,y)+\nu_-(x,y)]/2$. Note that because we have aligned the external magnetic bias field with the (111)-oriented NVs and there is no stray-field present, the magnetic field is completely aligned with (111)-oriented NVs and there is no contribution from an off-axis field $B_\perp$. The effective strain field is then $M_z(x,y)= D(x,y)- D_g$, where $D_g$ =2.87 GHz.

Due to a diamond's high index of refraction, significant NV PL undergoes total internal refraction and exits the diamond at the edges. Hence, edges are generally brighter and often saturate the corresponding camera pixels. Furthermore, PL emitted near the edges contains information from other parts of the diamond. Hence in Fig. 2, we filter out pixels within 1 μm of the edges as well as pixels with PL level exceeding a threshold value.

For magnetic microscopy (Fig. 4c), we look at the differential shift, $\Delta \nu = [\nu_+(x,y) - \nu_-(x,y)]/2$, which is then converted to magnetic field via $B_z = \Delta \nu / \gamma_e$. While the stray-field from a current-carrying wire has an off-axis component in general, the leading-order contribution from the off-axis field to $\nu_\pm$ is $\sim B_\perp^2$ and therefore contributes to the common-mode shift but not the differential shift.

## Section 4. Dark Matter Detection

One proposal for the directional detection of weakly interacting massive particles (WIMPs), a dark matter (DM) candidate, is via measurement of strain tracks in diamond left by WIMPs [8]. In this multi-stage proposal, a large number of diamond crystals of ~mm$^3$ volume are stored in a facility (e.g., deep underground) to interact with WIMPs that pass through the diamond detector (Fig. S1). Occasionally, a WIMP, or background particle such as a neutrino, initiates a nuclear recoil cascade via collision with a carbon nucleus and damages the diamond crystal lattice; the event is first registered and triangulated by photon, phonon, or charge-carrier collection (Fig. S1a). The diamond crystal in which the event is registered is isolated, and one then uses optical and/or x-ray techniques in to look for the associated strain feature, which needs to be eventually located with nm-resolution. The strategy would be to first perform-wide-field strain microscopy to locate the strain feature with μm or sub-μm precision (Fig. S1b), and at last perform microscopy with a super-resolution technique to achieve nm-resolution (Fig. S1c). To accomplish this goal, the diamond will need to be thinned down to avoid aberration; hence, the DMC we study here provides a representation of the eventual diamond sample for WIMP detection. An important feasibility question is whether the process of thinning down diamond may introduce strain features that mask the signal of a WIMP. Ref. [8] has shown that the expected fractional strain signal from WIMPs is $\sim 10^{-6}$; the standard deviation of strain measured in this work is $\sim 1\text{-}3\times 10^{-6}$, comparable to the state-of-the-art in Ref. [8] (before high-pass filtering) and the expected WIMP signal. Hence, our results show that it is possible to fabricate small diamond structure and be able to maintain low strain variation profile to retain sensitivity to WIMPs.

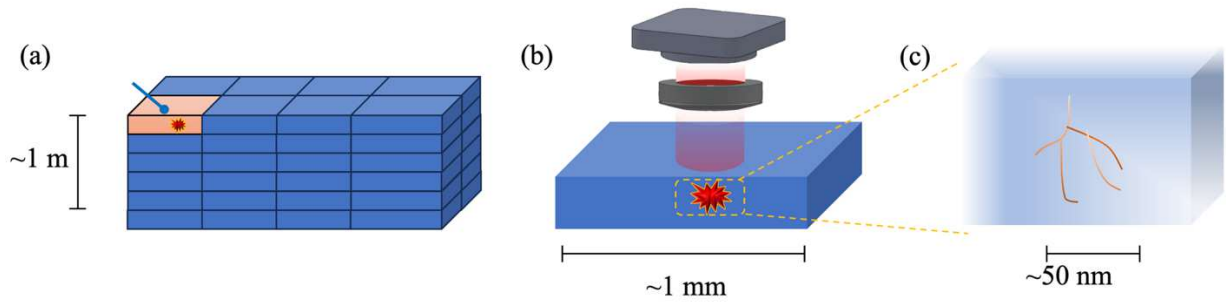

*Fig S1. Schematics of directional WIMP detection. (a) An event is registered and the particular ~mm³ volume of diamond crystal is identified for further measurement. (b) The registered diamond segment is taken for diffraction-limited microscopy, localizing the nuclear recoil track to a precision better than μm. (c) Finally, microscopy with ~nm resolution (e.g., via optical superresolution or x-ray techniques) is employed to measure the recoil track to nm precision. The incoming direction of the particle that created the nuclear recoil track can then be determined, potentially allowing discrimination of possible WIMPs from other background signals, such as from neutrinos.*

## Section 5. Spatial resolution

Spatial resolution in optical microscopy is given by the associated point spread function (PSF). The PSF is given by the Airy pattern which has the form $\text{PSF}_{\text{Airy}}(u)=(2J_1(u)/u)^2$, where $u$ is some dimensionless spatial scale and $J_1$ is the Bessel function of the first kind of order one. The first zero occurs at $u = 3.8317 \approx 4$; the length scale corresponding to this $u$ is the resolution according to the Rayleigh criterion. It is generally more convenient to use a Gaussian profile to approximate the PSF. The Airy pattern, to leading order in $u$, is $\text{PSF}_{\text{Airy}} = 1 - u^2/4 + O(u^4)$. A Gaussian profile of the form $\exp(-u^2/4)$ has the same expansion. Therefore, the following is a suitable Gaussian PSF:

$$\text{PSF}(x,y) = \frac{1}{\pi\sigma^2} e^{-(x^2+y^2)/\sigma^2}$$

and we identify $2\sigma$ as the spatial resolution.

In Fig. 4d, we fit the two neighboring strain features to a Gaussian profile. The underlying strain features (before being broadened by optical diffraction) may not necessarily be symmetrical. To account for such asymmetry, we fit each feature to the form

$$M_z(x,y) = Ae^{-(x^2/\sigma_x^2 + y^2/\sigma_y^2)}$$

The larger strain feature has a significant aspect ratio, while the difference in $\sigma_x$ and $\sigma_y$ is within the error bar for the small feature. We take the smaller of $\sigma_x$ and $\sigma_y$ of the small feature, and from it extract a spatial resolution of 450(30) nm.

The diffraction-limited resolution from the Rayleigh criterion expected for an imaging system is $\sigma=0.61\lambda/\text{NA}$, where $\lambda$ is the wavelength and NA is the numerical aperture of the objective. With a 0.9 NA objective and using λ=650 nm as the typical wavelength for NV PL, we have σ=440 nm, consistent with the experimentally extracted resolution.

**Section 6. Polymer-assisted dry-transfer**

The polymer-assisted DMC pickup and drop-off process is accomplished on a homemade deposition apparatus as shown in Fig. 1(b). With the diamond sample held on the translational stage on the left-hand-side, we first break a DMC off from the diamond with a sharp tip mounted on the motorized translational stage on the right-hand-side. The distance between the diamond and the sample surface is kept as small as possible by adjusting the tilt stage, which eliminates the chance of the DMC flipping upside down during the landing process. Fig. 5 (a)-(e) demonstrates the schematic of picking up a DMC and transferring it onto a target sample. A micro-structure of polymer made of 0.7 μm thick photoresist LOR5A/AZ-1512 is patterned on a transparent thin glass as shown in Fig. 5 (f). The polymer is located near the corner of the glass edges, which allows us to make sufficient contact of the polymer to the DMC by slightly tilting the glass substrate. A 50 μm DMC is successfully picked up by the polymer as recorded in Fig. 5(d)-(i), where the images focus on the DMC before and after it has been picked-up. Then the glass is moved to a location above the sample of interest. The sample shown in Fig. 5(j) is a patterned Hall bar structure of $RuO_2$ with four leads made of 2 nm Ti/100 nm Au. The DMC is vertically aligned to the sample location and gradually lowered down until it touches the Au leads as shown in Fig. 5(j)-(k). The tilt action and adhesion to the Au surface allows the release of the DMC from the polymer to the targeted sample in Fig. 5(k). Furthermore, with a subsequent operation shown in Fig. 5(l), we can also slightly manipulate the DMC to a desired location or orientation using a sharp tip.

Here we discuss differences between our technique and another recently reported dry-transfer technique using PMMA-PDMS stamp [9]. First, our technique does not require thermal-assistance, which is desirable for a temperature or air sensitive material, e.g,. van der Waals magnetic materials. Second, the technique in Ref [9] involves a PMMA layer that covers the whole diamond chip after the chip deposition. Thus, additionally opening a window in PMMA layer by electron-beam lithography is needed. In our work, the DMC is directly dropped off on the sample, and no additional fabrication process is needed. Lastly, since our technique does not have a layer of PMMA that holds down the chip, and thus we can recycle the chips.

We also note that Refs. [1,2] developed similar transfer technique for diamond membranes (thickness ~100 nm with lateral dimension ~100 μm). Difference between these works and our result are as follows: 1) we demonstrate the transfer of significantly smaller structures, with lateral dimension as small as 10 μm; and 2) our result does not require specialized surface preparation such as the use of polymer adhesion layer [1] (our technique can make use of it but does not require it), plasma treatment, or annealing [2], which can place limitations on the type of sample one can interface with a DMC.

As shown in Fig. 5, the use of a patterned polymer micro-structure enables us to pick up a specifically targeted chip without affecting other chips in the neighborhood. This enables a high-throughput and highly-efficient process for deploying DMCs. One can first detach many (or even all) DMCs onto a substrate (Fig. S2a). Then, each chip can be subsequently picked up for deposition (Fig. S2b). The advantage of this process is two-fold. First, detachment can be done just once, instead of having to be done each time a chip is to be positioned, enabling greater efficiency. Second, this allows one to have chips detached and stored on a substrate, instead of remaining attached on the mainframe. The latter is risky for storage and transportation because chips can be easily detached with accident in handling; and if a chip is not detached in a controlled manner, it most likely cannot be retrieved. Hence, detaching many chips on a substrate for storage allows one to safely store the chips for transportation and handling; and can reduce costs associated with having to fabricate DMCs again due to an accident.

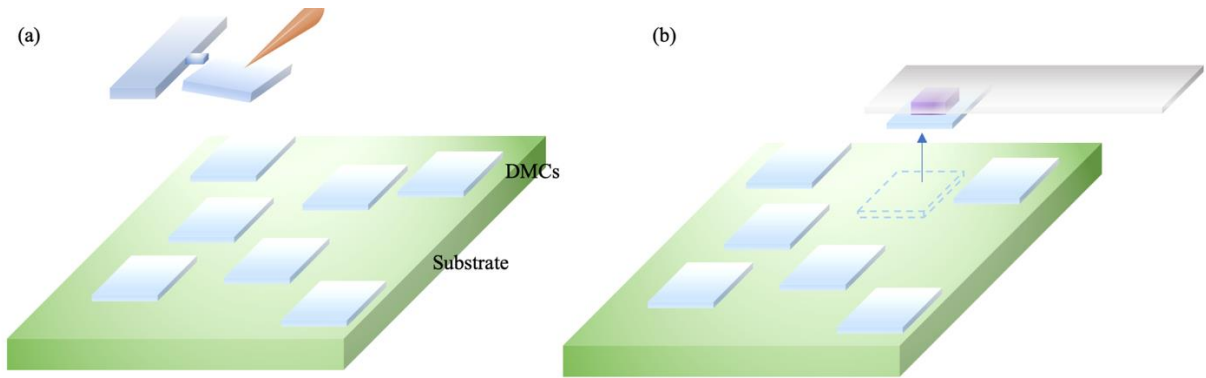

*Fig S2. Instead of detaching and then depositing a DMC one at a time, our approach allows one to take advantage of the targeted-pickup capability of the polymer-assisted transfer technique to enable a high-throughput transfer process. (a) One can first detach many DMCs onto a substrate, and subsequently (b) pick up an individual DMC to transfer onto the target sample.*